\documentclass[usenatbib,usegraphicx]{mn2e}

\usepackage{amssymb}
\usepackage{txfonts}
\newcommand{\aap}{A\&A}
\newcommand{\aaps}{A\&AS}

\newcommand{\aj}{AJ}
\newcommand{\apj}{ApJ}

\newcommand{\mnras}{MNRAS}
\newcommand{\pasp}{PASP}

\newcommand{\araa}{ARAA}

\newcommand{\dif}{\mathrm{d}}
\newcommand{\re}{r_\mathrm{e}}
\newcommand{\rd}{r_\mathrm{d}}
\newcommand{\Ie}{I_\mathrm{e}}

\begin{document}
\title[Incidence of the host galaxy]{Incidence of the
host galaxy on the measured optical linear polarization of blazars.}

\author[Andruchow et
al.]{I. Andruchow$^{1,2}$\thanks{andru@fcaglp.unlp.edu.ar}, Sergio
A. Cellone$^{1,2}$, Gustavo E. Romero$^{1,3}$\\
 $^1$Facultad de Ciencias Astron\'omicas y Geof\'\i sicas, Universidad
Nacional de La Plata, Paseo del Bosque, B1900FWA La Plata, Argentina\\
 $^2$Instituto de Astrof\'{\i}sica La Plata (IALP), CONICET--UNLP,
Argentina\\
 $^3$Inst.\ Argentino de Radioastronom\'{\i}a, C.C.5, 1894 Villa Elisa,
Buenos Aires, Argentina}

\maketitle

\begin{abstract}
We study the incidence of the underlying host galaxy light on the measured
optical linear polarization of blazars. Our methodology consists of the
implementation of simulated observations obtained under different atmospheric
conditions, which are characterised by the Gaussian $\sigma$ of 
the seeing function. The simulated host plus active nucleus systems span broad 
ranges in luminosity, structural properties, redshift, and polarization; 
this allows us to test the response of the results against each of 
these parameters.

Our simulations show that, as expected, the measured
polarization is always lower than the intrinsic value, due to the
contamination by non-polarized star light from the host. This effect is more
significant when the host is brighter than the active nucleus, and/or a
large photometric aperture is used. On the other hand, if seeing
changes along the observing time under certain particular
conditions, spurious microvariability could be obtained,
especially when using a small photometric aperture. We thus give
some recommendations in order to minimise both unwanted effects, as
well as basic guidelines to estimate a lower limit of the true (nuclear)
polarization.

As an example, we apply the results of our simulations to real polarimetric
observations, with high temporal resolution, of the blazar PKS\,0521$-$365.

\end{abstract}


\section{Introduction}
 \label{s_intro}

It is normally accepted that blazars are active galactic nuclei (AGNs) seen
with the line-of-sight very close to the axis of a relativistic jet
originated at its innermost regions \citep[e.g.][]{A93, UP95}. The radiation
of the jet is intrinsically polarized and relativistically boosted,
usually outshining the flux from any other component of the nucleus (e.g.,
accretion disk). Blazars present a strong and rapid variable flux \citep[e.g.,]
[and references therein]{RCCA02}, as well as high and variable
optical polarization \citep*[e.g.,][and references therein]{An05}.
This latter property is of particular interest, since its accurate
knowledge is important to correctly evaluate the intensity and
orientation of the magnetic field in blazars. Accurate variability studies are 
important because they allow to estimate the size of the emission region.

If it were possible to obtain detailed light curves at different 
wavelengths, identifying any correlation (or lack of it) between them, we 
could learn about the emission processes that produce the observed spectral 
energy distribution. From the point of view of optical linear
polarization variability, although several models try to explain 
its origin, the lack of good obsevational data is a problem 
that prevents against a satisfactory evaluation of the models.

An issue to be kept in mind in optical studies of AGNs, is the need to
separate the nuclear emission from the stellar light contribution of the
underlying host galaxy. This, in principle, is possible when the photometric
parameters of the host can be accurately measured
\citep[e.g.][]{KWJ04,N07,G08}.  However, this is usually a difficult task
for blazars, given their small angular sizes. A further complication is
added by the fact that any astronomical observation is affected by
systematic errors introduced by the instrument and (for ground-based
observations) the atmosphere. Whereas a quantitative and accurate knowledge
of these errors is always needed to obtain reliable data, in the case of
blazars that knowledge is imperative \citep[e.g.][]{CRA07}.

Several studies were made in the past to estimate the influence of
seeing on the parameters that describe the surface
brightness profiles of galaxies, for example, the effective radius
$\re$ \citep[e.g.][]{Sa93,Tr01}.
In general, these authors found that seeing scatters the light
from the centre of the galaxies to somewhat larger radii, with the
result that the observed mean values of the surface 
brightness are lower and effective radii are larger than their 
respective intrinsic values. In this way, it was shown that 
seeing affects directly the estimate of distances when a 
fundamental plane method is used, causing an overestimate of 
distances for distant galaxies.

Hence, we expect that the host galaxy light will affect polarization
measurements in blazars, and its effects will depend on the particular
atmospheric conditions (seeing) under which the observations are carried
out. This fact, which is already important for individual measurements,
becomes highly relevant for variability studies, because temporal changes in
the seeing conditions may lead to spurious variations in the blazar's
observed properties.

In this paper we study the effects of the host galaxy light on polarization
measurements of blazars, quantifying the dilution of the measured
polarization due to the host galaxy unpolarized light, as well as possible
spurious variations in the temporal polarization curves introduced by seeing
fluctuations. Our method relies on the analysis of simulated observations,
in the line of a previous study of seeing effects on photometric
microvariability observations of AGNs \citep*{CRC00}.

Section \ref{s_model} outlines the basics of our model, and
Sect.~\ref{s_genresul} gives a general view of the results. We then present
an application to real observations in Sect.~\ref{s-0521}. Our
conclusions are presented in Sect.~\ref{s_conclu}.

\section{Basics of the Model}
 \label{s_model}

The observed optical flux from an AGN can be considered, basically, as
shaped by two components: one non-polarized component coming from the host
galaxy, and other, with a certain amount of polarization, coming
from the active nucleus. From this point of view, the observed polarization
must be lower than the intrinsic polarization of the ``bare'' active
nucleus, due to the fact that the observed flux is a mix of those two
components. On the other hand, seeing affects the measurements; its
variations may introduce larger or smaller amounts of non-polarized light
from the host galaxy within the aperture used for the observations. Of
course, seeing also affects the light coming from the nucleus; however,
since the brightness distributions of the host galaxy and the nucleus are
different, any seeing variation will affect each component in different
proportions. Hence, the ratio between the (almost) totally non-polarized
flux from the host and the partially polarized flux from the active nucleus
will be affected by changes in the observing conditions, thus leading to
spurious variations when trying to measure the polarization behaviour of
AGNs against time.

In order to study how can the observed optical flux from AGNs be affected by
seeing variations, we performed simulations of observations as if they were
obtained under different conditions.
 For this purpose, we had to choose appropriate functions to describe the
 surface brightness profiles of the host galaxies, the brightness profiles
 of the active nuclei, and the atmosphere effects upon these profiles. We
 then generated a set of models of AGN + host galaxy systems spanning a
 suitably broad range in photometric and structural parameters, and
 convolved them with Gaussian functions representing different
 seeing conditions. Finally, we simulated polarimetric observations of these
 models, as if we were using a polarimeter with different
 apertures. Although our simulations are adjusted to the characteristics of
 the dual-beam polarimeter operating at CASLEO observatory, Argentina, which
 we used for our real observations \citep{An05}, our results should be quite
 general, and, in principle, they can be extended to other types of
 polarimeters. In the following subsections we describe these steps in
 detail.

\subsection{Active nucleus and host galaxy \label{s_agnhg}}

Our study is oriented to blazars, which, as a class, are the AGNs showing
the highest degrees of optical polarization. Since blazars are commonly
found in elliptical galaxies \citep[e.g.,][and reference therein]{N03, Sc00b}
the surface brightness profiles of their host galaxies can be described
by a de Vaucouleurs law \citep{DeV48}:
\begin{equation}
I_\mathrm{Gal}(r) = \Ie\,
e^{-7.67\,\big[\left(\frac{r}{\re}\right)^\frac{1}{4}-1\big]},
 \label{Ir}
\end{equation}
were $\re$ is the effective radius, and $I_\mathrm{e}$ is the effective
intensity (i.e., the value of the surface brightness where the radius is
$r=\re$). These are the only two free parameters in this equation, and they
determine the structure and magnitude of the host galaxy. We have always
considered hosts with circular isophotes, i.e. the profiles have azimuthal
symmetry.

Active nuclei, considered as structures isolated from the galaxies
hosting them, are point-like luminous sources \citep{KWJ04}. This
is true since the optical emitting regions in AGNs are typically unresolved
at extragalactic distances. So, a good approximation to represent the
brightness profiles of the simulated AGNs is a Dirac delta function. In
polar coordinates:
\begin{equation}
I_\mathrm{AGN}(r) = I_{0}\, \frac{\delta(r)}{\pi\mid r\mid}, \label{I_agn}
\end{equation}
were $I_{0}$ is the central intensity of the source, determining the
magnitude of the AGN.  The Dirac function is centred at the radial origin
of the host galaxy; this means, of course, that the active nucleus is
located at the centre of the system.

We need to give appropriate values to all the parameters involved
in equations (\ref{Ir}) and (\ref{I_agn}): $\re$, $I_\mathrm{e}$,
and $I_0$. Using the results from the surveys reported in
\citet{Sc00a}, \citet{Ur00} and \citet{Fa00}, where the
brightness values of several samples of blazars and their respective
host galaxies are studied, we chose values of $5$, $7.5$ and $10$
kpc as those spanning a representative range for $\re$.

Fluxes are proportional to $I_0$ and $I_\mathrm{e}$ for the AGN and the
host galaxy, respectively. As we want a fraction of polarized light in the final
expression, just the flux ratio is relevant, and thus we only need
to know the ratio $I_\mathrm{e}/I_0$. In order to set this ratio, we
must consider the magnitude difference, in a given waveband ($\Delta
m_{\lambda}$), between both components, and the expression relating it with
the corresponding flux ratio. The difference between host and nucleus total
apparent magnitudes is:
\begin{equation}
m_\mathrm{Gal} - m_\mathrm{AGN}= -2.5 \, \left[\log\left(22.67\, \re^{2}\,
\Ie\right) - \log\left(2\, I_0\right)\right] .
 \label{mag_difer}
\end{equation}

From this equation it is possible to derive an expression for the intensity
ratio as a function of the apparent magnitude difference,
$\Delta m_{\lambda}$. Both the host galaxy and the nucleus are at
the same distance from the earth; hence, the apparent magnitude diference is equal
to the absolute magnitude diference, $\Delta M_{\lambda}$, and so we work
with this latter value.  Again, using the values reported by \citet{Sc00a},
\citet{Ur00} and \citet{Fa00}, we choose as representative values:
$M_{\mathrm{AGN},V}$ = $-22$, $-24$, and $-26$, and the same values for
$M_{\mathrm{Gal},V}$. Differences of $\pm 4$\,mag are very infrequently
observed, so the only differences that we take into account are $\Delta M_V
= M_{\mathrm{Gal},V} - M_{\mathrm{AGN},V} = -2$, $0$, and $2$\,mag.

At this point, we have characterised a set of systems each consistent of a
host galaxy plus a nucleus. The next step is to put these configurations at
different distances, i.e., different redshifts $z$. We fixed five values for
$z$ at: $0.05$, $0.10$, $0.25$, $0.50$, and $1.00$. Adopting a cosmological
model with $H_0 = 70$ km\,s$^{-1}$\,Mpc$^{-1}$ and $q_0 = -\frac{1}{2}$, we
obtained the corresponding values of $\re$ in arcsec for the different
hosts.

Finally, we need to fix the polarization parameter $\alpha$, which
quantifies the intrinsic polarization of the nucleus. The theoretical upper
limit for synchrotron emission is $\alpha\sim 70\,\%$; however, this is not
a realistic value, because it relies on the assumption of a totally
homogeneous magnetic field, whereas in the more realistic case of a partially
inhomogeneous magnetic field, the observed polarization will be
substantially lower. The proposed values thus ranged from $1\,\%$ to
$50\,\%$, with varying steps: from $1\,\%$ to $5\,\%$ with a $1\,\%$ step,
from $10\,\%$ to $30\,\%$ with a $5\%$ step, and finally the value of
$50\,\%$. This represents a good sampling for the whole range of possible
intrinsic polarizations that we expect to find in blazars. Allowing
for dilution due to the host galaxy (see Sect.~\ref{s_genresul}), the
adopted range includes up to the highest observed polarization values in
blazars, about $\sim 45\,\%$ \citep{IBT82, MBB90}.

As it can be seen, there are many (495) combinations of all the parameters
giving different situations, each combination representing a different
model. Note, however, that a small subset of the combinations resulted in
indistinguishable models: those with the same value of $\Delta M_{V}$, the
same $\re$ in arcsec, and the same $\alpha$.

\subsection{Simulated observations \label{s_simulobs}}

Once defined the physical characteristics of the objects we are modelling,
the next step is to simulate the observations with a real instrument,
including the effects introduced by the atmosphere. So, we have now to
convolve the profiles of the AGNs and their respective host galaxies with
the seeing function, and then integrate these convolved functions within the
instrument aperture.

The image of a point source at the focal plane of a telescope is
described by the point spread function (PSF). For a medium- to
large-sized ground-based telescope with passive optics,
atmospheric seeing is the main contributor to the PSF. Seeing PSFs
can generally be well described by single Gaussians, Gaussians
with exponential wings, or Moffat functions. The most commonly
used function is the circular Gaussian:
\begin{equation}
\mathrm{PSF}(r) = \frac{1}{2\pi
\sigma^{2}}e^{-\frac{1}{2}\left(\frac{r}{\sigma}\right)^2}.
\label{psf}
\end{equation}
This is a simple function, characterised by just one parameter: the
dispersion $\sigma$. However, it describes appropriately the
effects of seeing upon the light from a point-like source. Besides
atmospheric effects, the PSF is also shaped by defects in the telescope's
optics, guiding and focusing errors, etc. These effects, in general, are not
azimuthally symmetric.  In the present simulation all these effects have not
been taken into account, because they would have complicated our modelling,
losing generality without any substantial gain in accuracy.  On the other
hand, they depend strongly on the particular characteristics of the
telescope and equipment used to obtain the measurements.  Hence, we opt for
a quite general approach, leaving any particular detail to be dealt with
more specific models to be developed by future researchers, should they find
it necessary.

Thus, we adopted the PSF given in Eq.~\ref{psf} to convolve the profiles
of the host galaxies and the active nuclei defined in Sect.~\ref{s_agnhg}.
In order to represent a wide range of seeing conditions, we considered
dispersions ranging from $0.25$ to $6$ arcsec, with a step of $0.25$
arcsec. These values correspond to full-width at half-maxima (FWHM) between
$\sim 0.6$ and 14 arcsec. We realize that the upper limit largely exceeds
what is expected for real observations; however, the adopted range is useful
to study the asymptotic behaviour of our results at both extremes.

All the functions depend only on $r$, which makes the calculation quite
simple. After the brightness profiles are convolved with the PSF, we have to
calculate the flux collected within an aperture for an instrument at the
focal plane of the telescope. Let $I^\mathrm{c}(\sigma, r)$ be the convolved
brightness distribution of a given source; the general expression for the
flux measured within an aperture of radius $\rd$ is then:
\begin{equation}
F^\mathrm{c}(\sigma, \rd)=\int_{0}^{\rd} \int_{0}^{2\pi}
I^\mathrm{c}(\sigma, r')\, r' \, \dif r' \dif \theta\, ,
\label{flujo}
\end{equation}
where $\theta$ is the azimuthal coordinate.  For the convolved brightness
distribution of the active nucleus it is possible to obtain the simple
analytical expression:
\begin{equation}
I^\mathrm{c}_\mathrm{AGN}(\sigma, r) = \frac{I_0}{\pi
\sigma^{2}} \, e^{-\frac{1}{2}\left(\frac{r}{\sigma}\right)^2}, \label{I_agn_c}
\end{equation}
which is just the Gaussian representing the PSF. Replacing this
expression within the integral in Eq.~\ref{flujo} we obtain the
observed flux from the AGN within the aperture:
\begin{equation}
F^\mathrm{c}_\mathrm{AGN}(\sigma, \rd) = 2\, I_0\, \left[1 -
  e^{-\frac{1}{2}\left(\frac{\rd}{\sigma}\right)^2}\right]\, .
\label{F_agn_c}
\end{equation}

For the host galaxy, there are several analytical methods that can be used
to compute the convolution of the brightness profile (a de Vaucouleurs law,
in this case) with the PSF. However, prioritising simplicity and saving of
computing time, we preferred a numerical approach, following the guidelines
given by \citet{Capa88}.  We thus implemented a \textsc{fortran} code to
obtain $I^\mathrm{c}_\mathrm{Gal}$ and $F^\mathrm{c}_\mathrm{Gal}$ by means
of numerical integration.

To characterise the instrument we took as a reference the CASPROF
photopolarimeter, currently used at CASLEO, San Juan, Argentina
\citep{FBV02, ACR03, An06}. This instrument uses as detectors a
pair of photo-multiplier tubes (PMTs), and has a similar design to
other dual-beam polarimeters operating at different observatories.
We set the radii of apertures at $\rd =2.8$, $5.6$, and $8.5$
arcsec, because these are the smallest apertures used with
CASPROF. Larger sizes are not expected to be used to observe
blazars.

The practical way in which all this was carried out, was just to
take into account for the calculations (either analytical or numerical ones)
the part of the function depending on $r$, and then including the
multiplicative constants, such as $I_0$ or $I_\mathrm{e}$.

For each model, we calculated then the fraction of polarized flux ($FP$),
i.e. the ratio between the polarized flux from the nucleus and the total
(AGN + host galaxy) flux, as a function of the seeing $\sigma$
(see Eq. \ref{psf}). This is the value of measured polarization 
expected for a fixed true polarization of the source 
($\alpha$). Any variation in $FP$ is thus only due to seeing variations and 
is spuriuos. The formal expression is:
\begin{equation}
FP(\sigma, \rd) = \frac{\alpha \, F^\mathrm{c}_\mathrm{AGN}(\sigma,
\rd)}{F^\mathrm{c}_\mathrm{AGN}(\sigma, \rd) +
F^\mathrm{c}_\mathrm{Gal}(\sigma, \rd)}. \label{FP}
\end{equation}
The fraction of polarized flux $FP(\sigma, \rd)$ was calculated for each
model and for the three aperture radii $\rd$. All this was carried out
using a \textsc{fortran} code, as mentioned previously.

\section{General Results}
 \label{s_genresul}

We have set the results from the simulations as plots of the polarization
fraction, $FP$, as a function of the seeing $\sigma$ and for each
aperture radius $\rd$. It is completely impractical to show results
for all the models, hence and in order to gain in clarity, we will discuss
the general results, emphasising particular results with special
interest. In the following subsections we will discuss the trends of the
results with the different variables of our models.

\subsection{Structural parameters}

\begin{figure}
\centering
\includegraphics[width=0.85\hsize]{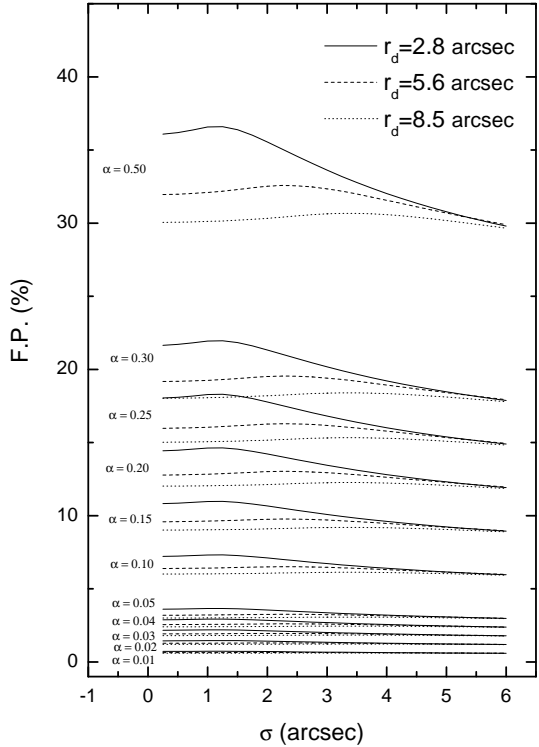}
\caption{Curves of the simulated fractional polarization
as a function of the seeing $\sigma$. Model parameters: $\re=5$\,kpc,
$\Delta M_V=0$, $z=0.05$. The three aperture sizes are shown for each
$\alpha$ (intrinsic polarization).}
\label{z005_a}
\end{figure}

\begin{figure}
\centering
\includegraphics[width=0.85\hsize]{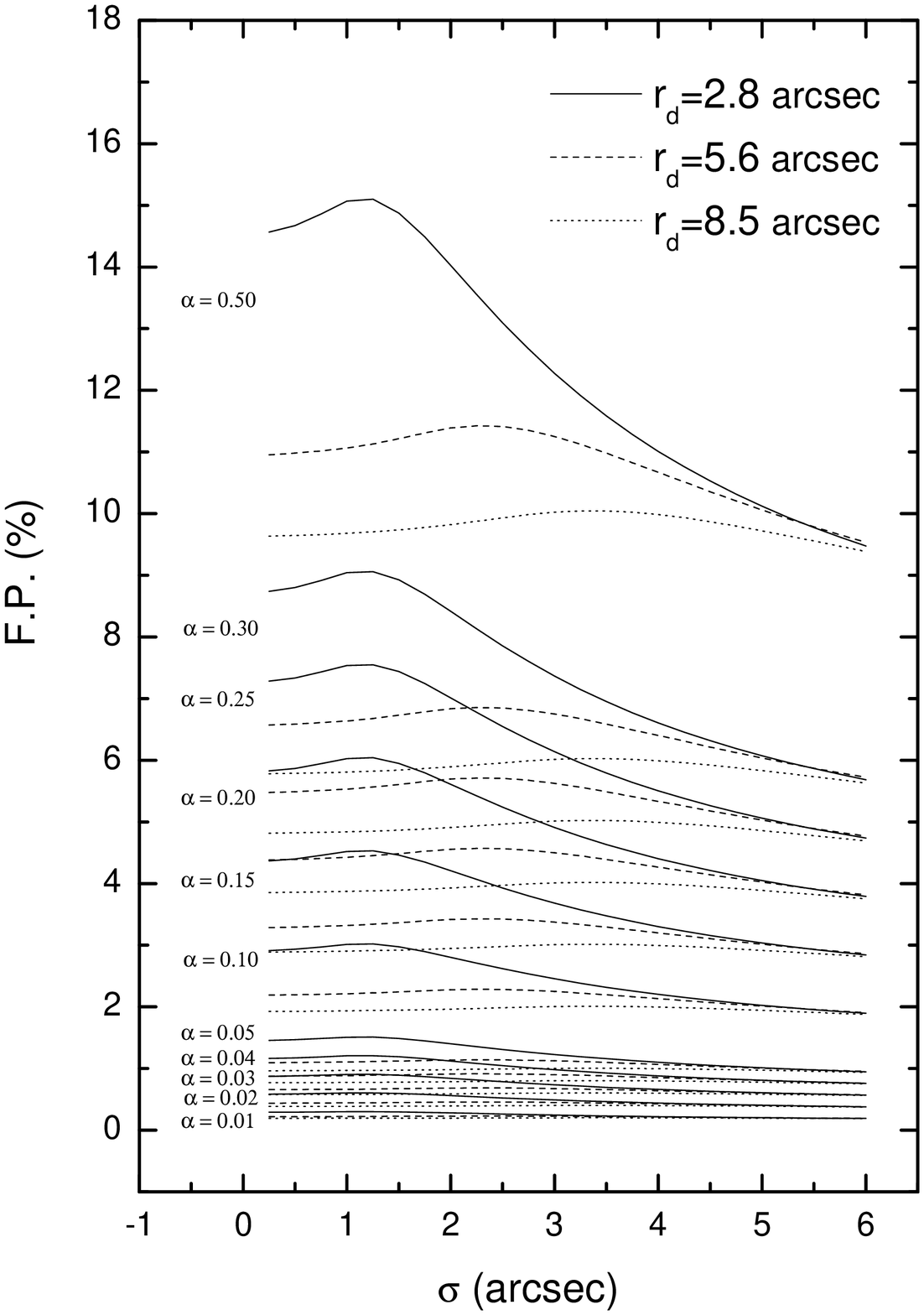}
\caption{Curves of the simulated fractional polarization
as a function of the seeing $\sigma$. Model parameters: $\re=5$\,kpc,
$\Delta M_V=-2$, $z=0.05$. The three aperture sizes are shown for each
$\alpha$ (intrinsic polarization).}
\label{z005_b}
\end{figure}

\begin{figure}
\centering
\includegraphics[width=0.85\hsize]{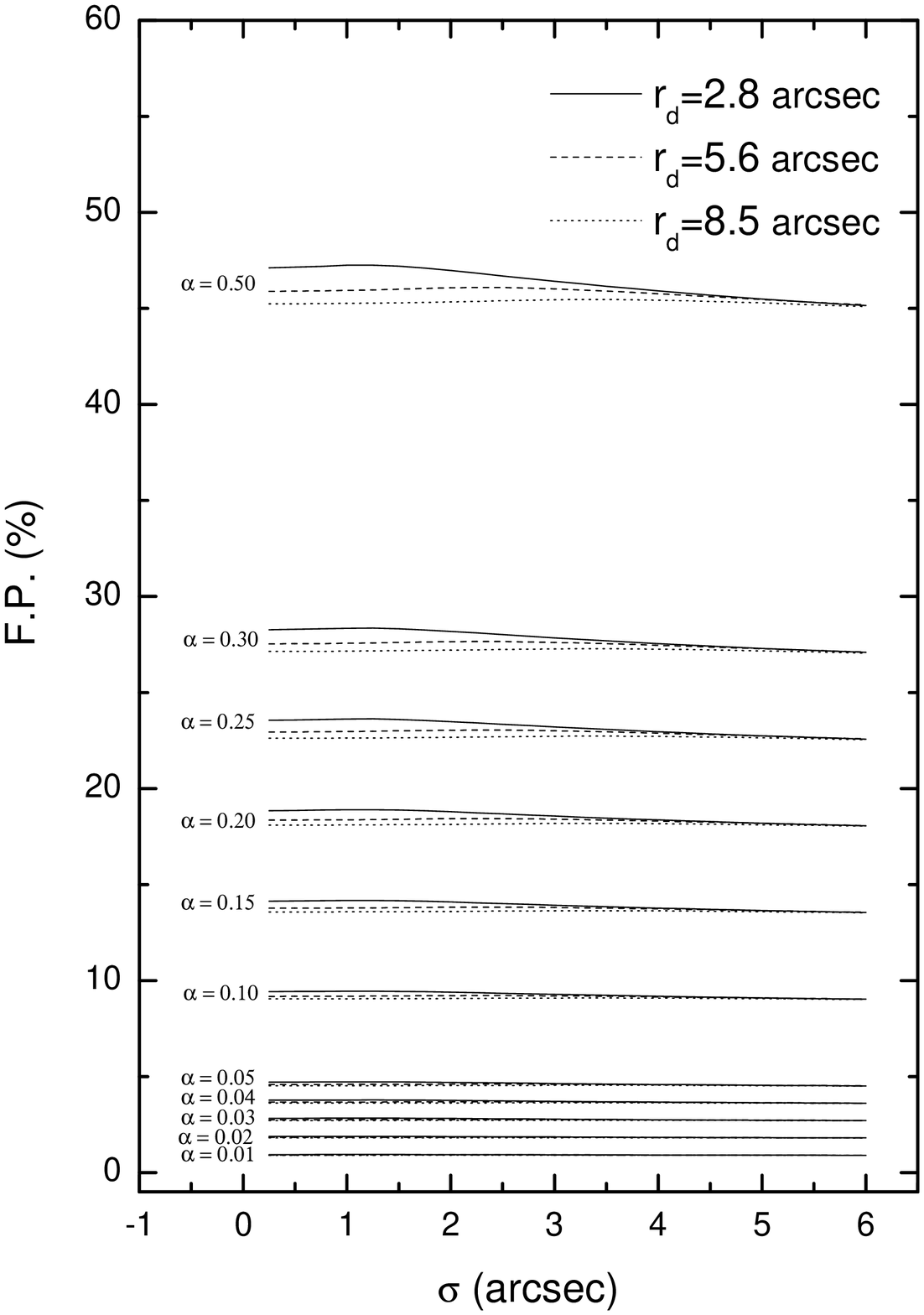}
\caption{Curves of the simulated fractional polarization as a
function of the seeing $\sigma$. Model parameters: $\re=5$\,kpc, $\Delta
M_V=2$, $z=0.05$. The three aperture sizes are shown for each
$\alpha$ (intrinsic polarization).}
\label{z005_c}
\end{figure}

Fig.~\ref{z005_a} shows results for the model with $\re=5$\,kpc, $\Delta
M_V=0$, and $z=0.05$. As it can be seen, the observed polarization, $FP$,
is always lower than the intrinsic polarization, $\alpha$. This is because
the polarized light from the nucleus is mixed with the unpolarized light
from the host galaxy \citep{N07}. For $\alpha=50\,\%$, the maximum
polarization fraction arriving at the top of the atmosphere is about
$36.4\,\%$ at $\sigma$ close to zero for the smallest aperture size
(according to our models). All the curves show this drop in the degree of
polarization, which depends strongly on the difference $\Delta M_V$, being
larger when the host galaxy is more luminous than the AGN, as evidenced by comparing
Figs.~\ref{z005_a}, \ref{z005_b}, and \ref{z005_c}.

Considering host galaxies with larger $\re$ and positive $\Delta M_V$
(nucleus brighter than host), the behaviour of the curves is qualitatively
similar, but with a much lower decrement of $FP$ with respect to $\alpha$.

\subsection{Observational conditions}
\label{ObsCond}
From Figs.~\ref{z005_a} to \ref{z005_c} one can see that all curves show a
maximum at $\sigma \approx \rd/2$, although this is more evident for smaller
$\rd$ and larger $\alpha$. This behaviour can be explained as follows.
At the lowest $\sigma$ values, the aperture contains almost all the
flux from the star-like nucleus, plus a part of the non-polarized light
from the more extended host galaxy. As $\sigma$ begins to increase, most 
of the polarized light from the nucleus is spread to an area still 
within the aperture, whereas a significant part of the unpolarized 
light from the host galaxy is spread out of the aperture; 
thus, $FP$ first grows. This happens until the seeing 
$\sigma$ is high enough that the fraction of (partially polarized) 
light removed from the nucleus becames larger than the fraction of (unpolarized) 
light removed from the host, thus lowering FP. The FP maximun is 
attained when the FWHM ($= \sigma / 0.4255$) is approximately similar 
to the aperture radius: when using larger apertures, a larger $\sigma$ 
has to be attained before it begins to spread out a significant fraction 
of polarized light from the nucleus.

The position of the FP maximun for a given aperture size 
has a very mild, if any, dependence with the host galaxy effective radius 
(at least, within the range of $r_\mathrm{e}$ used for the present 
simulations). It should depend, instead, on the slope of the convolved 
host galaxy surface brightness profile at the aperture edge, which 
determines the relative increment of galaxy light thrown off the aperture 
as $\sigma$ increases. This slope has only a mild dependence on 
$r_\mathrm{e}$. We verified these points using a set of artificial galaxy 
images with $r^\frac{1}{4}$ profiles convolved with a Gaussian PSF.

One important result from the observational conditions behaviour is that 
variations in the atmospheric conditions
could affect linear optical polarization measurements by introducing a spurious
variation component, thus undermining the reliability of microvariability
studies on a given source under certain (although rather extreme)
conditions. These would require, for example, a highly polarized AGN within a bright
host, observed under atmospheric conditions giving place to changes in the
seeing  from  $\sigma \simeq 2$~arcsec to more than $4$~arcsec. These
conditions would seem very unlikely to occur, but they are not impossible. In
general, telescopes used for the monitoring of AGN variability are not
large, modern instruments located at the best astronomical sites, because
these kind of studies demand large amounts of telescope time. On
the other hand, since astronomical polarimetry is a differential measuring
technique, it is usually assumed to be almost immune to mediocre atmospheric
conditions. Thus, it might not be so unlikely to face such an extreme
situation, with large seeing fluctuations along the observing time.

From our simulations, the changes in the polarization can be
divided into three regimes.  First, up to $\sigma \simeq 2$\,arcsec (FWHM
$\simeq 4.5$ arcsec) the change is small.  Between $\sigma \simeq 2$~arcsec
and $4.5$~arcsec ($4.5\,\textrm{arcsec} < \textrm{FWHM} <
10\,\textrm{arcsec}$) the change is higher and faster. From this point up to
the highest $\sigma$, the polarization behaviour flattens again. The second
regime is, thus, where the most significant spurious variability events
should be expected.

The amplitude of any possible spurious variation depends basically
on two parameters: the amplitude of the variation of the seeing function,
$\Delta \sigma$, and the aperture size. As it can be seen in
Figs.~\ref{z005_a} to \ref{z005_c}, smaller aperture sizes allow a
more accurate measurement of the AGN intrinsic polarization (by minimising
unpolarized light from the host), but, at the same time, they are the most
sensitive to changes in $\sigma$.

On the other hand, when $\sigma$ is high (regardless of whether it remains
constant or not), the fraction of polarized light diminishes.  This effect
is less noticeable for larger aperture radii, because the nucleus
polarization is already diluted by a large fraction of host light.
Although all models, with all possible combination of the different
parameters, do show this behaviour, the incidence of $\sigma$ variations
becomes almost insignificant when the nuclei are brighter than their
respective host galaxies (Fig.~\ref{z005_c}). The aperture with
radius $\rd = 5.6$\,arcsec always allows to get reasonably good quality data
while, at the same time, it minimises the effects due to changes in the
atmospheric conditions.

\subsection{Behaviour with redshift}
\begin{figure}
\centering
\includegraphics[width=0.85\hsize]{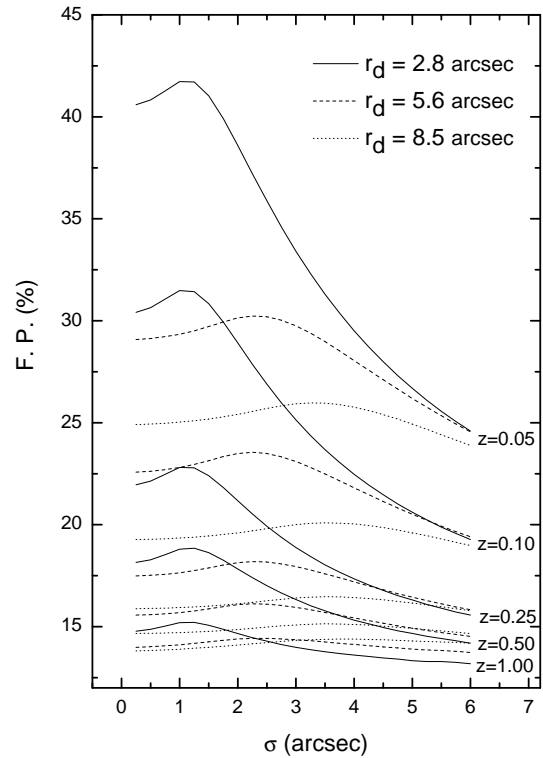}
\caption{Curves of the simulated fractional polarization as a
function of seeing. Model parameters:
$\re=10$~kpc, $\Delta M_{V}=-2$, $\alpha=1$. Three aperture
apertures are considered for each value of $z$.}
\label{todos_los_z}
\end{figure}

One interesting issue to study is what happens when considering the same
system host galaxy+nucleus at different redshifts. The modifications
introduced by changing $z$ imply only a geometrical effect, (for the
cosmogical model assumed, see Section~\ref{s_agnhg}). No evolution effects
were considered, and note that no additional correction for cosmological
dimming was applied, since the fluxes both from the host galaxy and from the
nucleus are equally affected by redshift, while K-correction effects can
safely be ignored for this study. In Fig.~\ref{todos_los_z}, curves for one
specific model ($\re=10$~kpc, $\Delta M_{V}=-2$) are shown. All curves take
only the value $\alpha=1$, which corresponds to a fully polarized
(i.e. physically unreal) nucleus. However, all the other possible curves
will behave in a qualitatively similar way, because $\alpha$ is a
multiplicative constant in the equation defining $FP$ (see
Eq.~\ref{FP}).

As $z$ increases, the effective radius of the host galaxy in
arcsec gets smaller, until most of the flux coming from the galaxy
is contained within the aperture. This leads to the following
effect: $FP$ gets lower with larger $z$ (at constant $\sigma$)
because of a larger unpolarized flux fraction from the host. At
the same time, variations of $FP$ with respect to $\sigma$ have
lower amplitudes for larger $z$.

\subsection{Temporal behaviour}
 \label{s_temporal}

So far, we have shown that the fractional polarization $FP$ is a
strong function of the seeing $\sigma$ under certain particular
conditions. We can now ask the question: how much do variations
in the atmospheric conditions affect the optical polarization
microvariability results for blazars?  To study this, we need a
seeing temporal behaviour curve. We propose that, by making
simultaneous observations of linear polarization and seeing, we
can estimate the influence of seeing on the actual polarization of
the source.

\begin{figure}
\centering
\includegraphics[width=0.9\hsize]{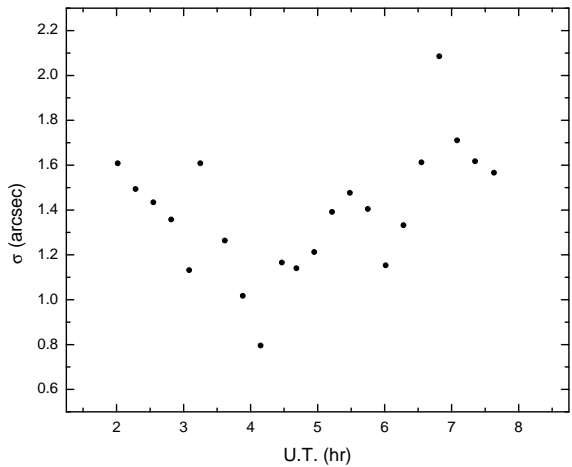}
\caption{Temporal evolution of $\sigma$ adopted from measurements made
during the monitoring campaign on the optical linear polarization of blazars
reported in \citet{An05}.}
\label{seeing_gral}
\end{figure}

In Fig.~\ref{seeing_gral} we present a temporal  curve for
$\sigma$ built from seeing measurements made at CASLEO during the
polarization campaign reported in \citet{An05}.
 We can then use this particular seeing curve to calculate the value of
$FP$ for each model. In order to do this, and for this particular
example, we ran all the simulations again with a $0.01$~arcsec step in
$\sigma$ (instead of $\Delta\sigma =0.25$~arcsec as in
Sec.~\ref{s_simulobs}) to have a better sampling. Then we matched each
value of $FP$ with its corresponding $\sigma$ for any given
time. In this way we obtained the temporal variation curves for $FP$ due
only to the changes in the atmospheric conditions. This procedure was carried
out for each of the three aperture sizes ($\rd = 2.8$, $5.6$, and
$8.5$ arcsec).

We found that the variations in atmospheric conditions translate into
variations in the calculated $FP$\, For the smallest aperture
size, the minimum and maximum values of $FP$ are matched with the maximum
and minimum on the seeing curve, respectively. This behaviour is hardly
distinguishable for the larger aperture sizes. As we already
pointed out, with the smallest $\rd$, the maximum amount of polarization is
recovered; however, seeing-induced changes can be large. For all
models, we found that the best balance between a high detection of
polarization and a low influence of seeing-induced variations was obtained
with an $\rd=5.6$~arcsec aperture.

Although all models show the same general (qualitative) behaviour, the
influence of seeing variations on $FP$ depends on the parameters
describing each model. Again, we got that the spurious effects due to seeing
variations become negligible when the nucleus is brighter than the host
galaxy. And this effect is further enhanced for large $z$, when the 
host is not only dim but also of small angular size, compared to $\rd$.

\section{An Example: Application to PKS\,0521$-$365 }
 \label{s-0521}

As an example of the results presented in previous sections, we applied the
method to the blazar PKS\,0521$-$365. We chose this particular source
because it is a well-studied blazar, which has a bright elliptical host
galaxy with well-known parameters. Regretably, bad weather conditions
prevented us from obtaining additional data of this object under different
observational setups (i.e., using different apertures, etc.), and so we had
to rely on observations from our monitoring program on the optical linear
polarization of blazars \citep{An05}.  Although this situation does not
allow a full testing of our simulations, we judge that the example we
present is at least illustrative for our purposes.

\begin{figure}
\includegraphics[width=0.49\hsize]{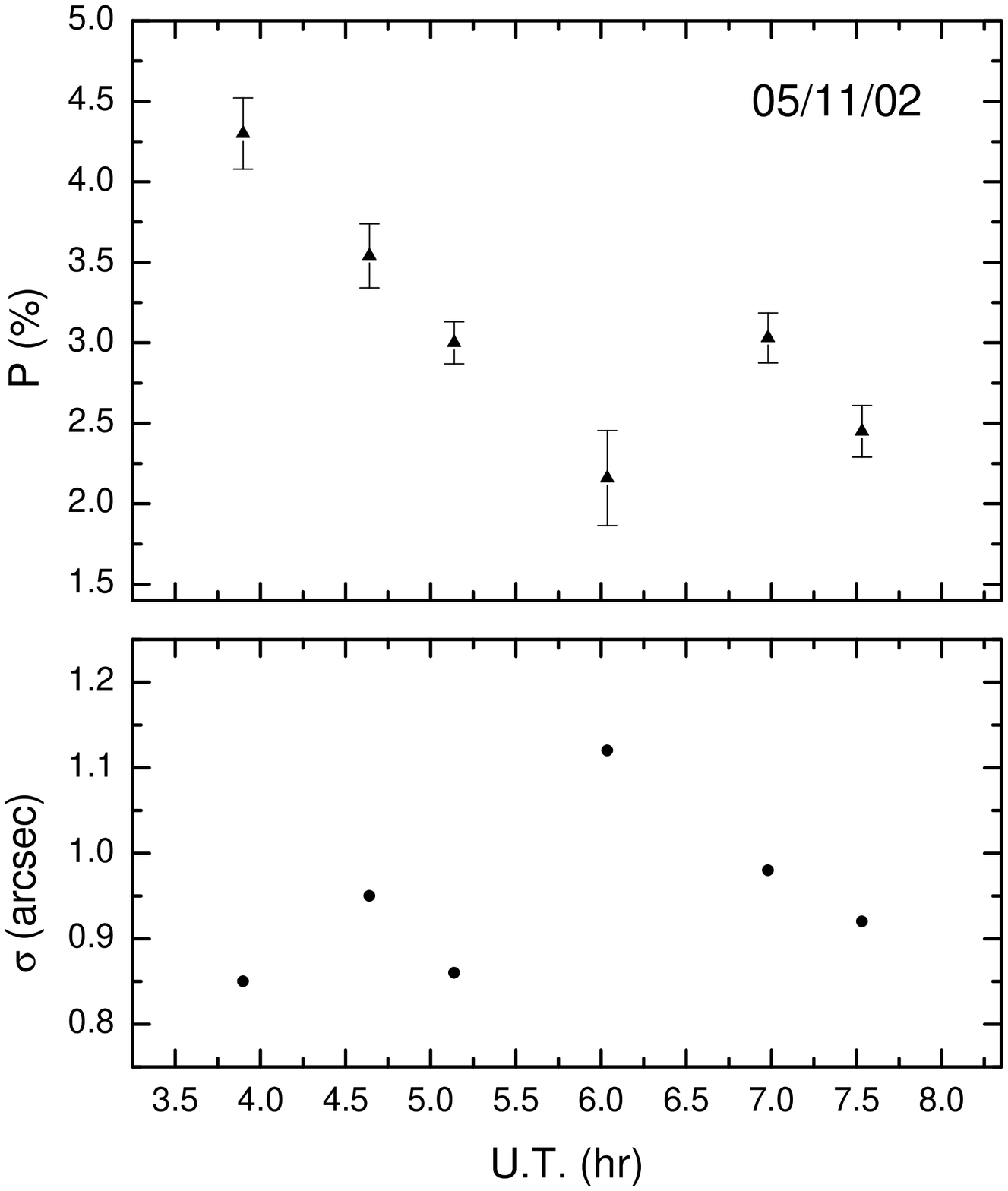} \hfill
\includegraphics[width=0.49\hsize]{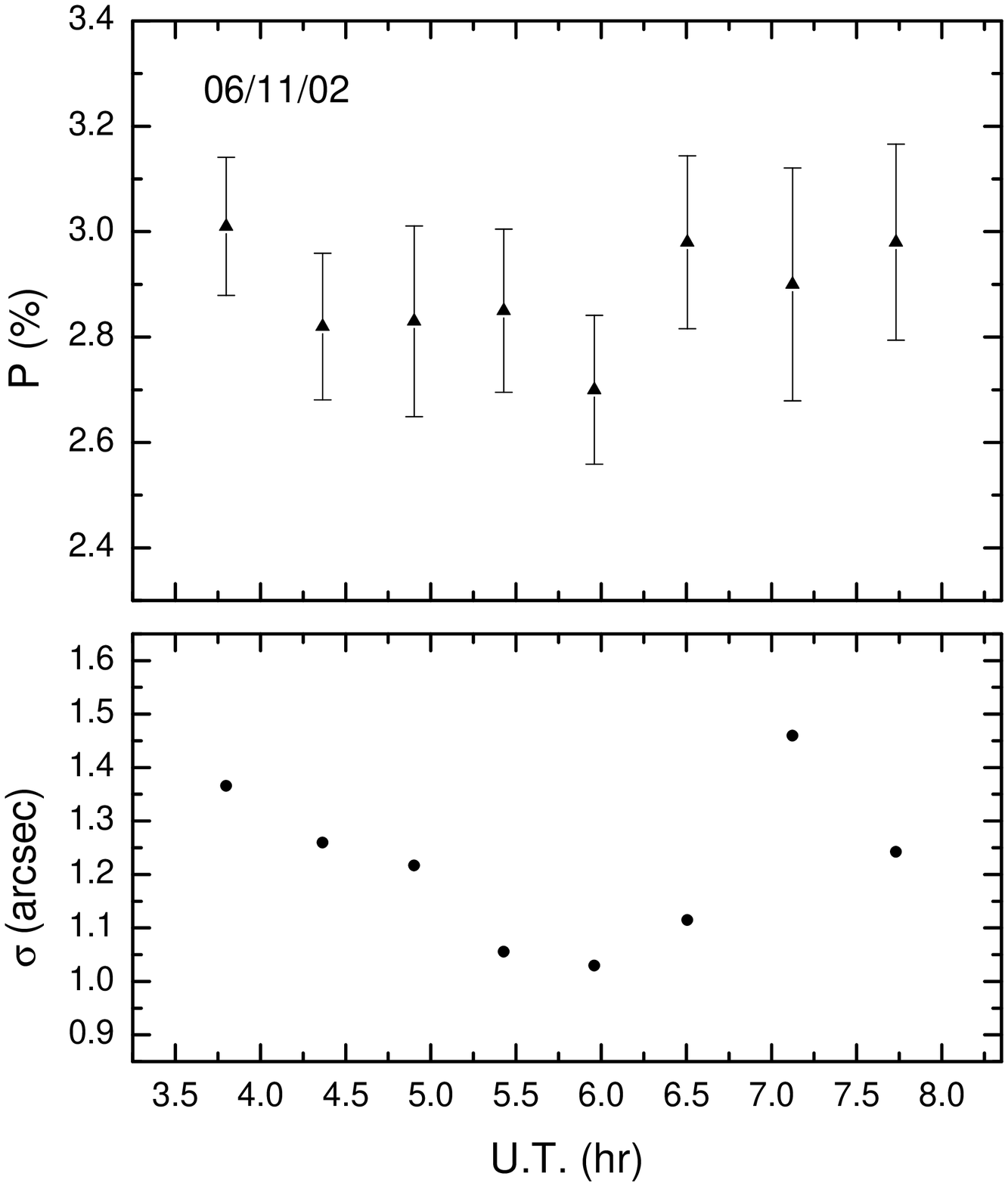}
\caption{Temporal behaviour of the degree of polarization, $P$, for
PKS\,0521$-$365 (upper panels) and the simultaneous seeing measurements,
$\sigma$ (lower panels). Left column: observations for the night of Nov.\
05, 2002; right column: similar curves for the night of Nov. 06, 2002.}
\label{polar_seeing_obser}
\end{figure}

The observations were carried out using the 2.15-m Jorge Sahade telescope at
CASLEO, San Juan, Argentina, during two consecutive nights on November 2002.
The data were collected using the CASPROF photopolarimeter, in an on-off
regime, i.e., performing a target observation followed by a near-sky one to
allow for the corresponding subtraction of the sky polarization
contribution. A few points (affected by poor transparency during the night)
were removed after a first analysis. Off-target observations before and
after the target pointing were interpolated to increase accuracy in the
subtraction procedure.  The seeing measurements were made with a DIMM-type
monitor placed close to the dome. In Fig.~\ref{polar_seeing_obser} we show
the observational results for the two nights along which we followed the
source. The error bars are calculated as in \citet{MBR84}, from
photon-noise statistics. Significant polarimetric microvariability can be
seen, at least for the first night.

\subsection{Specific Model}
 \label{SpecMod}

In order to evaluate which fraction (if any) of the linear
polarization variability observed for PKS\,0521$-365$ is due to
seeing fluctuations, we must now choose the specific model which
best represents the object's structural properties. Using the
results of the surveys reported in \citet{Sc00a}, \citet{Ur00} and
\citet{Fa00}, the host galaxy and nucleus parameters that we
adopted for PKS\,0521$-365$ were:
\begin{itemize}
    \item  $m_{\mathrm{HOST},\, R}=14.60 \pm 0.01$ mag,
    \item  $m_{\mathrm{AGN},\, R}=15.28 \pm 0.10$ mag,
    \item  $\re=2.80 \pm 0.07$ arcsec,
    \item  $z=0.055$.
\end{itemize}

\begin{figure}
\centering
\includegraphics[width=0.85\hsize]{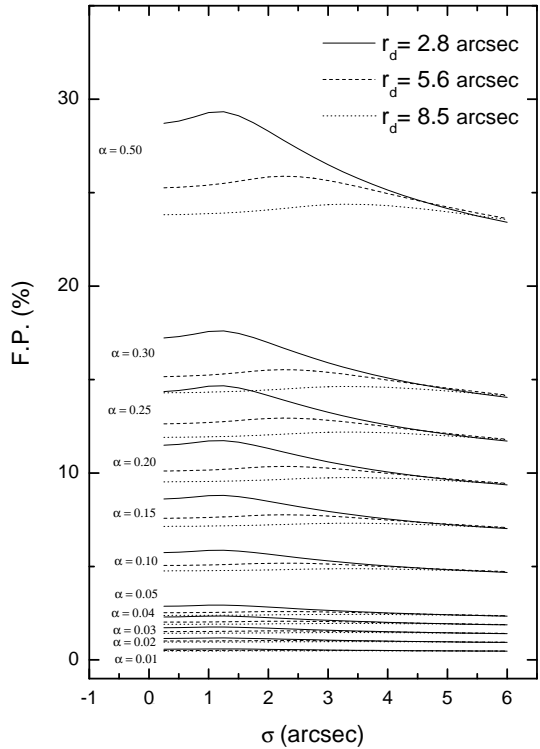}
\caption{Simulated fractional polarization as a function of the
seeing, $\sigma$, for the particular case of the source
PKS\,0521$-$365. Parameters of the model: $\re=2.8$~arcsec, $\Delta
M_{V}=-0.34$, $z=0.055$. Three aperture values are shown for each $\alpha$.}
\label{0521_mod}
\end{figure}

The apparent magnitudes are given in the $R$-band of the Johnson-Cousins
system. Using the values of the colour index, $V-R$, for the host
galaxy and the nuclear point like source from \citet{Ur00}
we obtained the $V$-band apparent magnitudes. With these values, the
magnitude difference gave $\Delta m_{V}=\Delta M_{V}=-0.34$. The plots for
each aperture size and for all values of $\alpha$ for the
corresponding model are shown in Fig.~\ref{0521_mod}.

In Table~\ref{0521_obs} we present the corresponding observational
results for the two nights in which we followed the source in Nov.\ 2002. Column
1 gives the date; the number of points for each night is given
in column 2; column 3 gives the mean polarization values; column 4 shows the
respective standard deviations; column 5 is the time difference between the
maximum and minimum $P$ values; and column 6 is the variability result,
``V'' for variable and ``NV'' for non variable. From the observations, the
mean value for the degree of polarization is about $3\,\%$. Using the
results from the simulation corresponding to the adopted model, we looked
for the value of $\alpha$ which best reproduces $FP=3\,\%$ when $\sigma$
is close to zero for an $\rd=5.6$ arcsec aperture (see also
Fig.~\ref{0521_mod}); in this way we adopted $\alpha=0.06$ as the
theoretical value according to the models.

This $\alpha=0.06$ (i.e., 6 percent) represents then the intrinsic
polarization of the active nucleus. For each night, we took the
corresponding seeing measurements and, assuming that the degree of intrinsic
polarization was always constant at $6\,\%$, we obtained the behaviour of
$FP$ vs.\ time from the results of the simulations for the
chosen aperture.

\begin{figure}
\includegraphics[width=0.49\hsize]{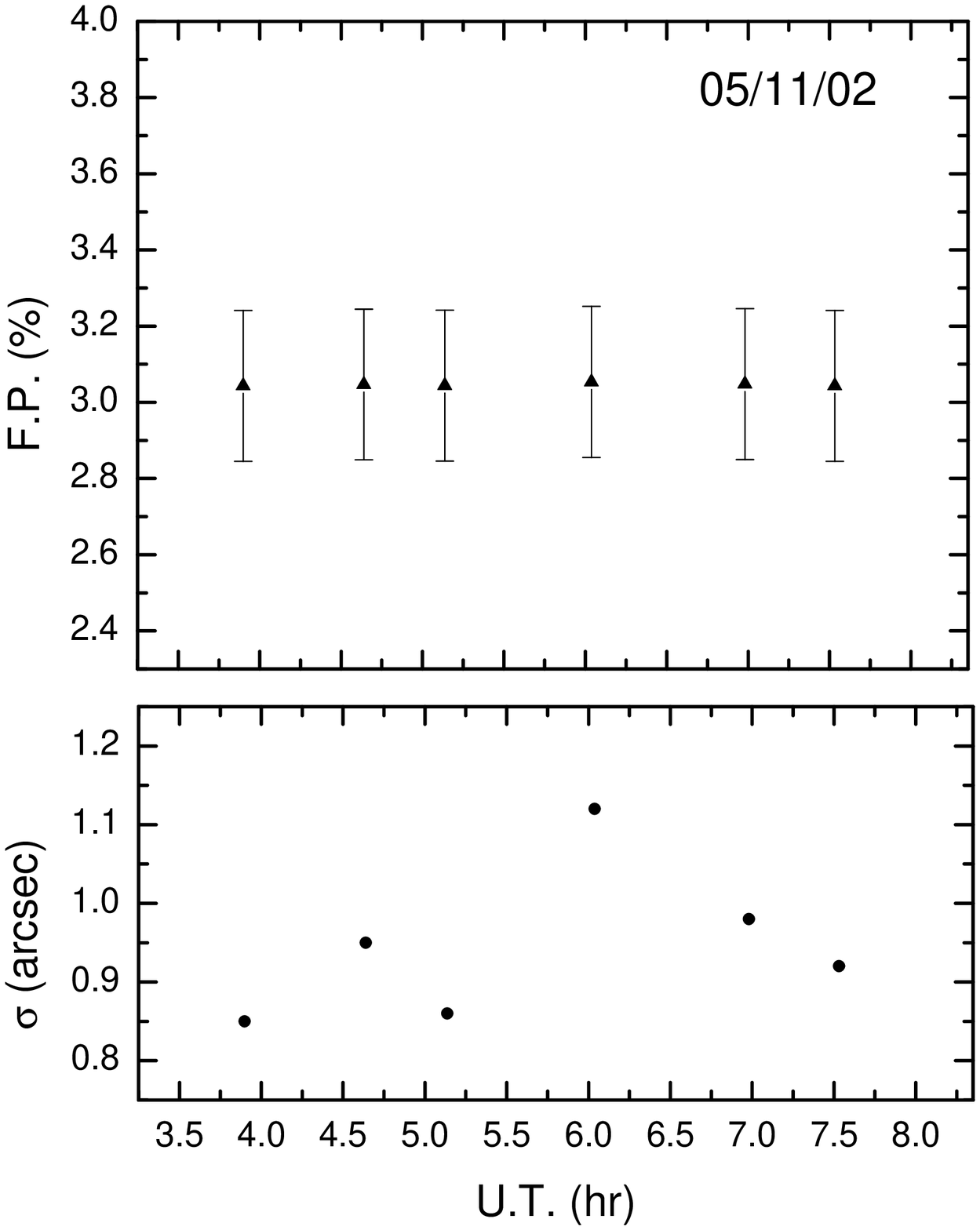} \hfill
\includegraphics[width=0.49\hsize]{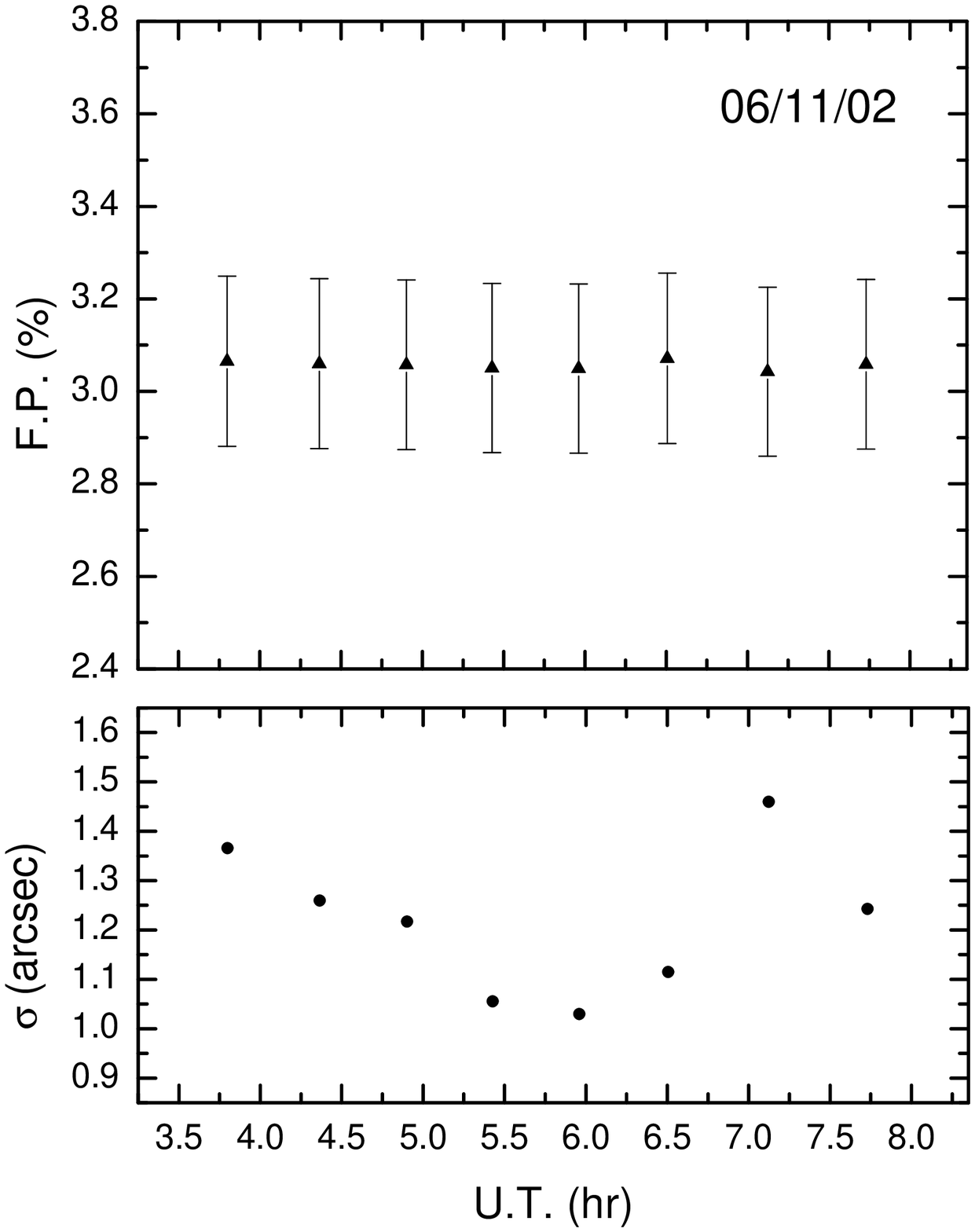}
\caption{Upper panels: Variation curve of the simulated $FP$ due
only to seeing changes as a function of time, for the nights of
Nov.\ 5, 2002 (left) and Nov.\ 6, 2002 (right).  Lower panels:
behaviour of $\sigma$ from simultaneous observations. Left column:
Nov.\ 5, 2002. Right column: Nov.\ 6, 2002.} \label{f_simulacion}
\end{figure}

The variations thus obtained for $FP$ should represent those due only to
the fluctuations in the atmospheric conditions during the observing
session. In Fig.~\ref{f_simulacion} we show the results: upper panels
correspond to the results of the simulations whereas lower panels show the
seeing behaviour for each night. The error bars of each point were
estimated from the actual observational errors. The average
error of each individual observation point was about $5\%$
, so, we adopted this average observational error as the error for each
simulated data point. Although the fluctuations have very small
amplitudes, the general trend was recovered: this means that the maximum of
$FP$ corresponds to the minimum of the seeing and vice versa.

\begin{table}
\caption{Statistical results of the degree of polarization
microvariability for the blazar PKS\,0521$-$365. Observations. 
Here, $n$ stands for the number of observations, $V/NV$ for ``Variable'' or 
``Not Variable'', and the remaining symbols have the usual meaning (see text).}
\label{0521_obs}
\begin{center}
\begin{tabular}{cccccc}
\hline\hline
Date & $n$ & $\langle P \rangle$ & $(\mathrm{rms})_P$ & $\Delta t$ &
V/NV\\
$[$d/m/y$]$ & & [$\%$] & & [h] & \\
\hline
 05/11/2002 & 6 & 3.05 & 0.769 & 2.1387 & V \\
 06/11/2002 & 8 & 2.88 & 0.105 & 2.1621 & NV \\
 \hline
\end{tabular}
\medskip
\end{center}
\end{table}

\begin{table}
\caption{Statistical results of seeing measurements for each
night.} \label{seeing_est}
\begin{center}
\begin{tabular}{ccccc}
\hline\hline
Date & $\langle \sigma \rangle$ & $(\mathrm{rms})_\sigma$ & $\Delta \sigma$ &
$\Delta t$ \\ %
 $[$d/m/y$]$ & [arcsec] & & & [h]  \\
 \hline
 05/11/2002 & 0.95 & 0.10 & 0.27 & 2.1387 \\
 06/11/2002 & 1.22 & 0.15 & 0.43 & 1.1627 \\
 \hline
\end{tabular}
\medskip
\end{center}
\end{table}

\begin{table}
\caption{Statistical results of the polarization fraction flux
microvariability for the blazar PKS\,0521$-$365. Simulations. 
Meaning of the symbols as in Table~\ref{0521_obs}.}
\label{0521_sim}
\begin{center}
\begin{tabular}{cccccc} \hline
\hline \noalign{\smallskip} Date & $n$ & $\langle FP \rangle$ &
$(\mathrm{rms})_{FP}$ & $\Delta t$ & V/NV \\
${\rm [d/m/y]}$ & & [$\%$] & & [h] & \\
\hline
05/11/2002 & 6 & 3.05 & 0.004 &2.1387 & NV \\
06/11/2002 & 8 & 3.06 & 0.005 & 0.6174 & NV \\
\hline
\end{tabular}
\medskip
\end{center}
\end{table}

\subsection{Statistical Results}
 \label{s_resul}

The way we used to compare  observed and  simulated values was
through the statistical analysis of each data set. In
Table~\ref{seeing_est} we present the results of the statistical
analysis of the seeing measurements. Column 1 is the date; column 2 is
the mean value of seeing $\sigma$; column
3 is its standard deviation; column 4 is the amplitude of $\sigma$
variations; and column 5 is the time difference between the maximum and
minimum $\sigma$. In Table~\ref{0521_sim} we present the
corresponding statistical results for $FP$. The colums have similar
meanings to those in Table~\ref{0521_obs}.

The ratio between the dispersion of the observed polarization and the
dispersion of the simulated polarization fraction can be used as a
quantitative test to assess whether or not spurious (i.e. seeing-induced)
variations are significant. In this sense, we propose that, if
$(\mathrm{rms})_{P}/(\mathrm{rms})_{FP} > 1$ the changes in the night
condition do not affect the variability results.  Otherwise, if
$(\mathrm{rms})_{P}/(\mathrm{rms})_{FP} < 1$ the variability result could
be modified by a seeing time curve variation. For the case of
PKS\,0521$-$365, on both nights, this ratio was
$(\mathrm{rms})_{P}/(\mathrm{rms})_{FP}\gg 1$; hence, from a statistical
point of view, there was no significant spurious variability due to seeing
changes in the observed polarization during any of the nights.
There is a rather worrisome hint for a broad trend between seeing
and polarization time curves during the first night; however, $\Delta P$ is
much larger than expected just from seeing variations. We can thus be
confident that the variations detected for the source had an intrinsic
origin in the blazar.

During the second night, the seeing values were higher and changes
occurred with a higher amplitude than during the first night (see
Table~\ref{seeing_est}). However, this relatively high amplitude
would have not been enough to introduce any spurious variation
component by itself. On the other hand, the high seeing values
during the whole observational session had direct influence on the
data quality. In extreme cases, the larger error bars could have
masked possible low amplitude variations.

Probably because of the relatively small amount of polarization shown by
PKS\,0521$-$365 and to the particular observational conditions during both
nights, we were not able to obtain any firm conclusions regarding the influence
of changes in the atmospheric conditions on the polarimetric variability
results for this particular blazar. In any case, we want to point out
that this example shows that the
metodology is actually aplicable to a real case. Further studies, involving
enough data to improve the statistics, and using different aperture radii,
are needed in order to obtain more general results.

\subsection{Inference of the intrinsic linear optical polarization}
 \label{s_cota}

As it is pointed out by \citet {N07}, it is difficult to have an
estimate of the true optical polarization of blazar nuclei.
One implication of the approach that we present here is that it allows
to estimate a lower limit to the intrinsic value of the active nucleus
polarization. As we pointed out in the case of PKS\,0521$-$365 observations,
by collecting both polarization and seeing data, and knowing the host/AGN
photometric parameters and the redshift of the object, we can apply the
model results as corrective terms, obtaining an estimation for the true
polarization for any given measurement. This estimation is interpretated 
as a lower limit because, under the assumptions made for our models, at least is 
needed to have that amount of polarization at the sources in order to 
reproduce the observational data.

Defining $FF$ as the fraction of the total flux originated in the
active nucleus which we measure within a given aperture, the
relationship between $FF$ and the value of the observed 
polarization, $P$ (wich may variate both by seeing and intrinsic causes), is given
by:
\begin{equation}
P = \alpha \, FF\, . \label{P-FF}
\end{equation}

The flux fraction $FF$ is thus the right-hand member of Eq.~\ref{FP}
divided by $\alpha$, and is obtained from our models. With this value and the
observed polarization ($P$) we can recover the intrinsic AGN polarization
($\alpha$).

Despite of the fact that we find no practical
way to provide future observers with a detailed output from our
models, serving as ``ready-to-use'' corrections to their measurements, we
can nevertheless give them a few numbers which can serve as a guide to
estimate a lower limit of the true (nuclear) polarization. This is done in
Table~\ref{tabFF}, which should be read as a double-entry table with the
redshift in column 1, and gives the flux fractions corresponding to the
maxima in Figs.~\ref{z005_a}-\ref{todos_los_z}. Column 2 corresponds to an
aperture radius $r_{d}=2.8$\,arcsec, column 3 to $r_{d}=5.6$\,arcsec, and
column 4 to $r_{d}=8.5$\,arcsec. These maxima correspond approximately to
seeing values $\sigma = 1.0$\,arcsec, $\sigma = 2.2$\,arcsec, and $\sigma =
3.2$\,arcsec, respectively for each aperture radius. Note that the positions
of these maxima, as said
in Sect.~\ref{ObsCond}, do not depend on the hosts effective radii (at least
for the range in $r_\mathrm{e}$ that we used), so we just give our results
for the three different magnitude differences ($\Delta M_V$) considered in
Sect.~\ref{s_model}.

We can now go one step further and, by applying the above described process
to each individual data point in an observing session (if we are studying a
time series), we can follow the behaviour of $\alpha$ during any given
observing session. As an example, we applied these ideas to the blazar
PKS\,0521$-$365 observations. Notice that we are proposing a temporal
dependence for $FF$ and $P$. So, Eq.~(\ref{P-FF}) is now re-written as:
\begin{equation}
P(t) = \alpha(t) \, FF(t)\, . \label{P-FPbis}
\end{equation}
The temporal dependence in $FF$ (and hence in $FP$) is through the seeing
$\sigma$ variations. However, the temporal behaviour of the observed
polarization $P$ will be due to these seeing-induced variations plus any
variability intrinsic to the source. Thus, knowing the temporal behaviour of
$FF(t)$ from the models and the seeing measurements, it is in principle
possible to recover the actual polarimetric behaviour of the active nucleus,
$\alpha$(t), from the observed polarimetric curve $P(t)$.

\begin{figure}
\centering
\includegraphics[width=0.49\hsize]{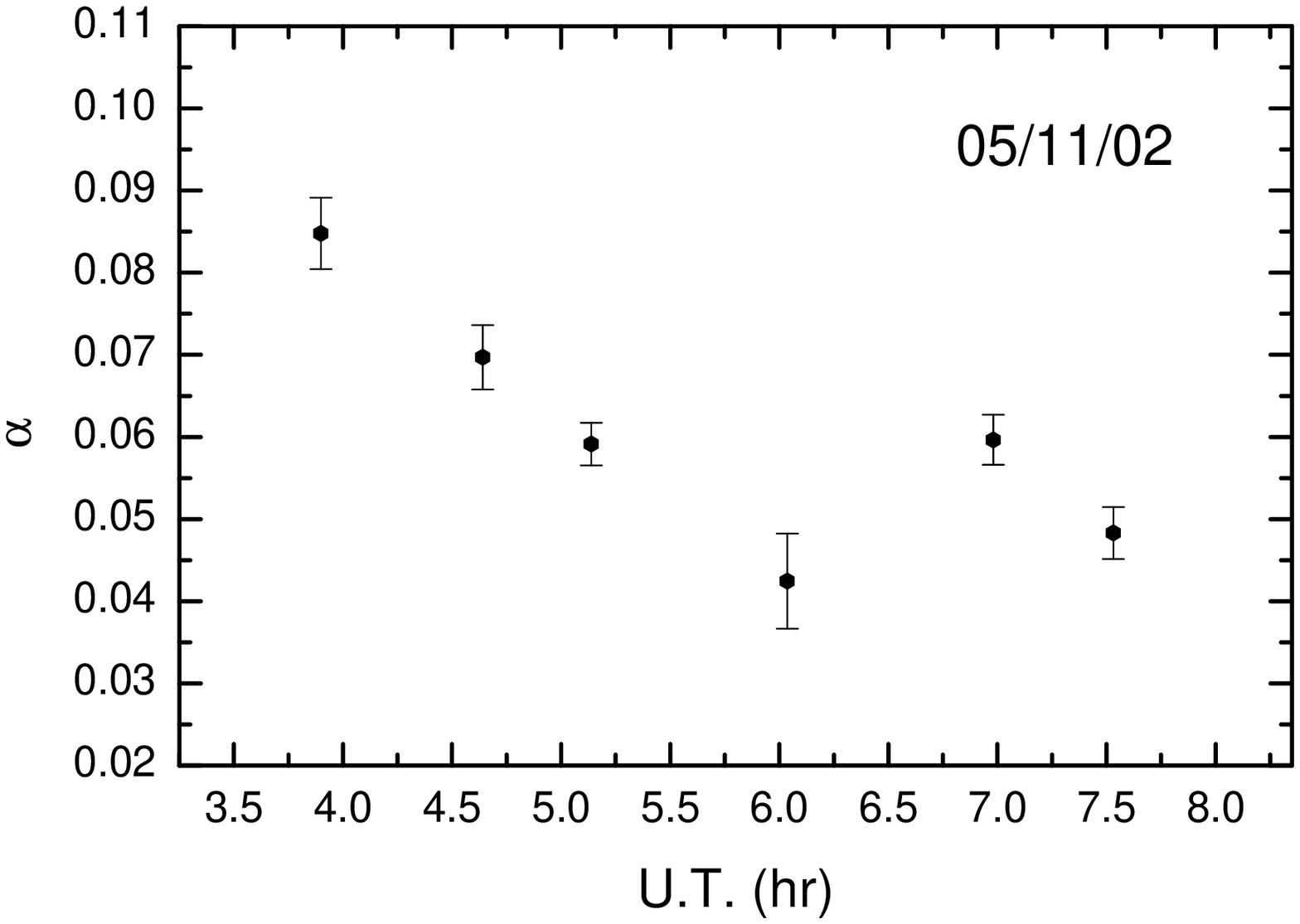}\hfill
\includegraphics[width=0.49\hsize]{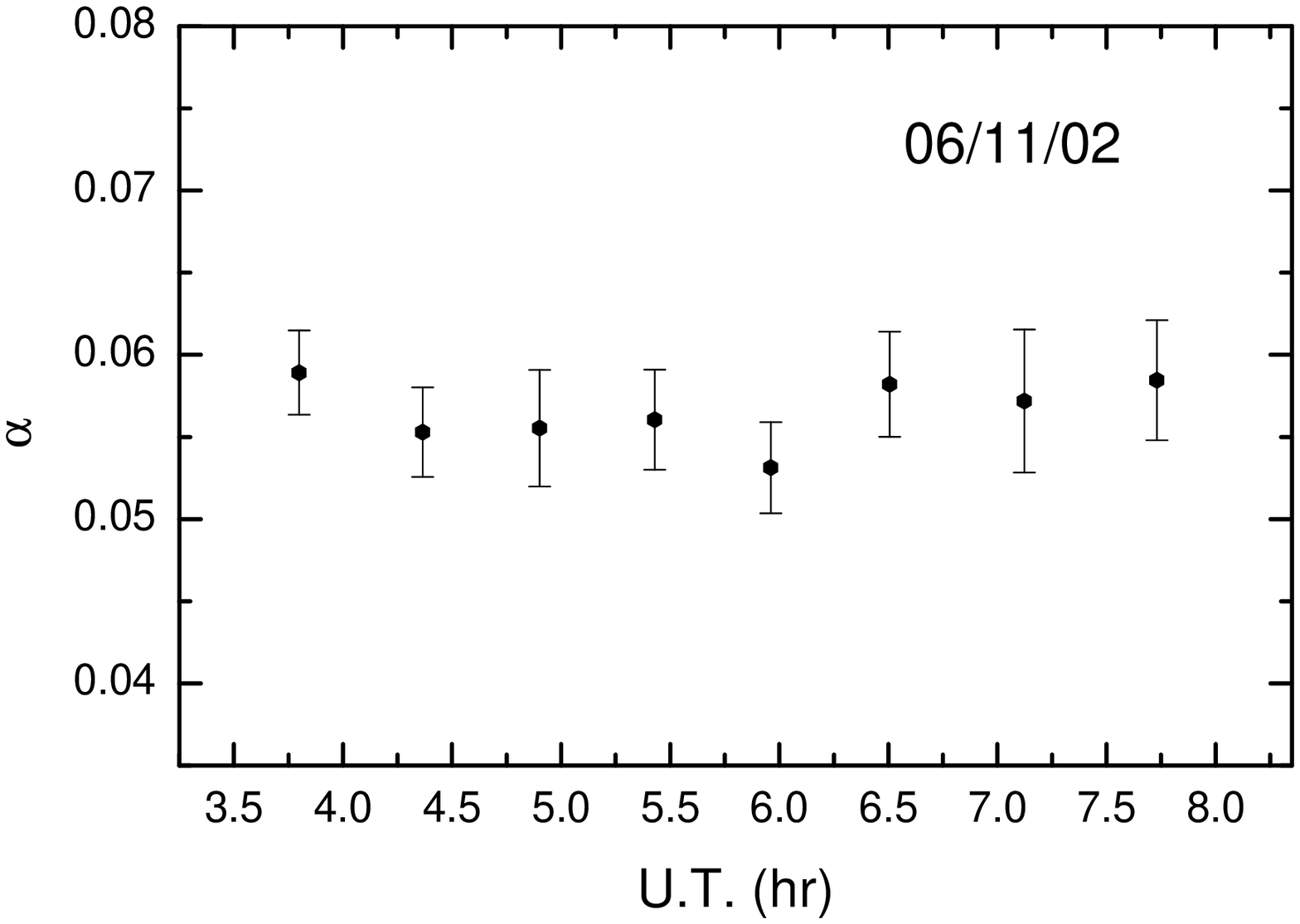}
\caption{Variability curves for the $\alpha$ 
parameter in the case of PKS\,0521$-$365.} \label{P_FP_alfa}
\end{figure}

Applying this to the case of PKS\,0521$-$365, we calculated $\alpha$ for each
observational instant at the nights of November 5 and 6, 2002. In
Fig.~\ref{P_FP_alfa} we present the temporal variation curves for $\alpha$.

During both nights, since the amplitude of the $FP$ curves was low, the
general trend of the curves resulted similar for $P$ and $\alpha$ (see
Figs.~\ref{polar_seeing_obser} and \ref{f_simulacion}, upper pannels, for
the behaviour of $P$ and $FP$ during the night).  These facts can be used
as a confirmation of the variability results obtained from the
observations. However, in order to have a clearer picture of the different
systematics affecting optical polarization measurements in blazars, it would
be necessary to test our ideas with other targets. Nonetheless, we mention these
first steps because we believe that they may
contain a relevant potential for studying what is actually happening at the
regions were the optical polarized radiation is generated.

\begin{table}
\caption{Maximum values of $FF$ for the models in
  Figs.~\ref{z005_a}-\ref{todos_los_z}.}
\label{tabFF}
\begin{center}
\begin{tabular}{c|ccc} \hline
\hline \noalign{\smallskip}
%$z$/$r_{d}$ & $2.8 [\sigma = 1]$  & $5.6 [\sigma = 2.25]$ & $8.5 [\sigma =
%  3.25]$ \\
%           & arcsec & arcsec & arcsec \\
%
$z$  & $FF_1$ & $FF_2$ & $FF_3$ \\
\hline
&  \multicolumn{3}{c}{$\Delta M_V = -2$} \\
\cline{2-4}
0.05 & 30.20 & 22.85 & 20.09 \\
0.10 & 23.49 & 18.51 & 16.88 \\
0.25 & 18.17 & 15.81 & 14.97 \\
0.50 & 16.10 & 14.77 & 14.44 \\
\cline{2-4}
&  \multicolumn{3}{c}{$\Delta M_V = 0$} \\
\cline{2-4}
0.05 & 73.19 & 65.15 & 61.33 \\
0.10 & 65.96 & 58.91 & 56.16 \\
0.25 & 58.36 & 54.24 & 52.63 \\
0.50 & 54.78 & 52.24 & 51.56 \\
\cline{2-4}
&  \multicolumn{3}{c}{$\Delta M_V = +2$} \\
\cline{2-4}
0.05 & 94.51 & 92.18 & 90.92 \\
0.10 & 92.44 & 90.05 & 88.99 \\
0.25 & 89.84 & 88.21 & 87.51 \\
0.50 & 88.43 & 87.34 & 87.04 \\
\hline
\end{tabular}
\medskip
\end{center}
\end{table}

\section{Conclusions}
 \label{s_conclu}

We have modelled the incidence of the host galaxy on optical linear
polarization measurements for blazars. We show that, knowing the relevant
photometric parameters of the host galaxy (effective radius, effective
surface brightness, magnitude difference with respect to the active nucleus)
and the value of the seeing $\sigma$, an estimate of the intrinsic 
value of the optical polarization can be obtained. This value is always 
higher than the observed polarization.

Moreover, if the degree of polarization presented by the blazar is
high enough (how high is ``high enough'' depends on the system nucleus +
host galaxy) and seeing time variations do occur (under conditions
corresponding to the second regime mentioned in Section~\ref{ObsCond}), a
spurious component in the measured polarization curve may result,
especially for nearby blazars with relatively bright hosts.
So, in general, if the seeing remains stable during the night, the most
suitable aperture will be a small one, in order to minimize the
underestimate of the polarization. On the other hand, if seeing is poor
and variable, we found that an intermediate-sized aperture (in our
case $\rd = 5.6$\, arcsec) may give a good compromise between spurious
variations obtained with smaller radii and a severe subestimation of the
intrinsic polarization obtained with large-sized apertures.

 In principle, these spurious fluctuations may be removed from the observed
polarization curve, provided that the seeing temporal evolution along the
night is known. Simultaneous measurements with a seeing monitor are needed
in the case of polarimetric observations done with an instrument like the
one used for the present work. CCD polarimetry, on the other hand, has the
advantage that the PSF (including instrumental effects besides seeing) can
be directly measured on each science frame. However, care must be taken in
this case since different PSFs are usually obtained for the ordinary and
extraordinary images. Whereas the general conclusions of our work may still
apply for CCD polarimetry, a particular modelling will probably be needed in
this case.

\medskip

\noindent{Acknowledgements:} This project was supported by the
Argentine Agencies CONICET (GRANT PIP\,5375) and ANPCyT (GRANT
PICT\,03-13291 BID\,1728/OC-AR). G. E. Romero thanks support for 
the National Natural Science Foundation of China (GRANT 10633010). 
The authors thank CASLEO staff, and they wish to
dedicate this paper to the memory of Rebeca Morales; CASLEO will not be the
same without her. The authors want to thank the anonymous referee for valuable 
comments that really helped to improve the present paper.

\end{document}